\documentclass[
 aps,
 prl,
 reprint,
 superscriptaddress,
 longbibliography
]{revtex4-2}

\usepackage{amsmath,amssymb,amsfonts}
\usepackage{bm}
\usepackage{mathrsfs}
\usepackage{mathtools}
\usepackage{graphicx}
\usepackage{xcolor}
\usepackage{xurl}
\usepackage{hyperref}
\usepackage{etoolbox}

\AtBeginEnvironment{thebibliography}{%
  \sloppy
  \emergencystretch=2em
}

\hypersetup{
 colorlinks=true,
 linkcolor=blue,
 citecolor=blue,
 urlcolor=blue
}

\newcommand{\nc}{\newcommand}
\nc{\rnc}{\renewcommand}
\nc{\nn}{\nonumber}

\nc{\del}{\partial}

\rnc{\Im}{\mathrm{Im}\,}
\rnc{\Re}{\mathrm{Re}\,}

\nc{\bra}{\langle}
\nc{\ket}{\rangle}

\nc{\tcr}{\textcolor{red}}
\nc{\tcb}{\textcolor{blue}}

\begin{document}

\title{Hidden Dyson Universality in Inverse-Spectral Geometry}

\author{Momo Hayashi}
\email{2223088@ed.tus.ac.jp}
\affiliation{
Department of Physics, Tokyo University of Science,
Kagurazaka 1-3, Shinjuku-ku, Tokyo 162-8601, Japan
}

\author{Kazumitsu Sakai}
\email{k.sakai@rs.tus.ac.jp}
\affiliation{
Department of Physics, Tokyo University of Science,
Kagurazaka 1-3, Shinjuku-ku, Tokyo 162-8601, Japan
}

\date{\today}

\begin{abstract}
Dyson universality typically manifests itself in local eigenvalue statistics.
Here we show that its signature survives a nonlinear inverse-spectral
reconstruction and reappears in the matrix geometry of the reconstructed
operator. Using a dressing transformation, we map each unfolded spectrum
to a deformation $f(x)$ of a fixed harmonic oscillator and represent it in
the common oscillator basis by $F_{mn}=\bra m|f|n\ket$. We resolve the
matrix-element weight into shells of fixed distance $d=|m-n|$, corresponding
to the energy-transfer channels of the reference oscillator, and characterize
the resulting distribution by distance-shell moments. Independently calibrated
on Gaussian $\beta$-ensembles, these moments vary smoothly with $\beta$ and
distinguish the GOE, GUE, and GSE. With this calibration fixed, applying the
same diagnostic to the nontrivial zeros of the Riemann zeta function places
the reconstructed operators in the GUE sector. Thus, the GUE character of the
zeros is recovered not through direct statistics of the input levels, but from
the distance-resolved geometry of the reconstructed operator.
\end{abstract}

\maketitle
\emph{Introduction.---}
Random-matrix universality is most sharply expressed in the local statistics
of unfolded spectra.  Unfolding rescales the spectrum to unit local mean
spacing, removing the smooth variation of the level density.  For the
Gaussian ensembles, the corresponding symmetry classes are the orthogonal,
unitary, and symplectic ensembles, denoted GOE, GUE, and GSE, respectively
\cite{Dyson1962,MehtaBook,ForresterBook}.  Dyson class is conventionally
diagnosed through level spacings, spacing ratios, two-point correlations,
spectral form factors, and related eigenvalue statistics
\cite{AtasEtAl2013}.  These diagnostics act directly on the eigenvalue
sequence.  A natural question is whether the same universality also has an
operator-level manifestation.  Recent studies of operator growth and
Krylov-space dynamics provide one route to this question
\cite{ParkerEtAl2019,BalasubramanianEtAl2022,CamargoEtAl2025}, but they
normally start from a given Hamiltonian.  Here we ask the inverse question:
if only a spectrum is given, can its Dyson universality survive a nonlinear
inverse-spectral reconstruction and reappear in the geometry of the
reconstructed operator?

The nontrivial zeros of the Riemann zeta function provide a particularly
sharp setting for this question.  Montgomery's pair-correlation conjecture
predicts GUE correlations for the unfolded zeros at large height, and
Odlyzko's computations provided strong numerical evidence
\cite{Montgomery1973,Odlyzko1987,Odlyzko2001}.  The Hilbert--P\'olya idea,
on the other hand, asks whether the zeros arise as the spectrum of a
self-adjoint operator~\cite{BerryKeating1999,Connes1999}.  A complementary
inverse-spectral route is to reconstruct an operator from the prescribed
zeros.  Wu and Sprung found fractal structure in a semiclassical
Schr\"odinger potential associated with the zeros~\cite{WuSprung1993}, and
a nonlinear top-down dressing transformation introduced by Ramani,
Grammaticos, and Caurier~\cite{RamaniEtAl1995} was later applied directly
to them~\cite{vanZylHutchinson2003,SchumayerEtAl2008,
SchumayerHutchinson2011}.  What has remained unclear is whether the Dyson
class itself survives such a nonlinear reconstruction and reappears as a
simple structure of the reconstructed operator.

Here we show that it does.  We apply the same dressing transformation to
every input spectrum, reconstruct a deformation $f(x)$ of a fixed harmonic
oscillator, and represent it in the common oscillator basis by
$F_{mn}=\bra m|f|n\ket$.  The distance $d=|m-n|$ is the magnitude of a
Bohr frequency of the reference oscillator and defines common
energy-transfer shells.  Distance-shell moments measure how the
matrix-element weight is distributed over these shells and thereby
characterize the distance-resolved geometry of the reconstructed operator.
The nontrivial point is not that the reconstructed operator encodes the
prescribed spectrum, but that a small set of these basis-resolved observables
recovers its Dyson-class information.  Calibrated independently on Gaussian
$\beta$-ensembles, the moments vary smoothly with $\beta$ and distinguish
the GOE, GUE, and GSE.  With this calibration fixed, and without adjustment
to the zeta data, the operators reconstructed from the Riemann zeros fall in
the GUE sector.  Their residual deviation from GUE decreases with height and
is concentrated predominantly in the low-frequency shells.

\emph{Inverse-spectral construction.---}
For each input spectrum, we construct a deformation of a fixed reference
harmonic oscillator.  We take
\begin{equation}
 H_0
 :=
 -\frac{d^2}{dx^2}
 +
 \frac{x^2}{4}.
\end{equation}
Its normalized eigenstates $|n\ket$,
$n\in\mathbb{Z}_{\ge 0}$, satisfy
\begin{equation}
 H_0|n\ket=E_n|n\ket,
 \qquad
 E_n=n+\frac12 .
\end{equation}

Let $s$ denote the spectral source.  We use $s=\beta$ for a Gaussian
$\beta$-ensemble and $s=\zeta$ for the Riemann zeros, suppressing
individual realization or window labels.  The cases $\beta=1,2,4$
correspond to the GOE, GUE, and GSE, respectively.  For the Riemann-zero
data, the dependence on the window height is indicated explicitly by $T$. We denote the corresponding unfolded
target levels by
\begin{equation}
 \varepsilon_0^{(s)}
 <
 \varepsilon_1^{(s)}
 <
 \cdots
 <
 \varepsilon_{N_{\rm lev}-1}^{(s)} ,
\end{equation}
with unit mean spacing.  Here $N_{\rm lev}$ is the number of target
levels retained in the dressing construction.

Following \cite{RamaniEtAl1995,vanZylHutchinson2003}, we apply the same
top-down dressing transformation to every input spectrum.  It produces a
deformation $f^{(s)}(x)$ such that the low-lying spectrum of
\begin{equation}
 H^{(s)}:=H_0+f^{(s)}(x)
\end{equation}
reproduces the target levels
$\{\varepsilon_n^{(s)}\}_{n=0}^{N_{\rm lev}-1}$.  The dressing recursion
and the restoration of the energy origin are described in the End Matter.

Since the same reference operator $H_0$ is used for every input spectrum,
its eigenbasis provides a common harmonic-oscillator reference frame.
We therefore project the deformation onto the first $N_{\rm b}$ oscillator
states and denote the resulting finite matrix by $F^{(s)}$, with elements
\begin{equation}
 F_{mn}^{(s)}
 :=
 \bra m|f^{(s)}|n\ket
 =
 \int_{-\infty}^{\infty}
 \phi_m(x)f^{(s)}(x)\phi_n(x)\, dx.
 \label{eq:Fmn}
\end{equation}
Here $0\le m,n<N_{\rm b}$, and
$\phi_m(x):=\bra x|m\ket$ is the normalized $m$th eigenfunction of
$H_0$.  The symmetric dressing prescription gives
$f^{(s)}(-x)=f^{(s)}(x)$ and hence
$F_{mn}^{(s)}=0$ for odd $|m-n|$.

The index separation has a direct dynamical meaning.  Introducing the
reference Liouvillian
\begin{equation}
 \mathcal L_0:=\operatorname{ad}_{H_0},
 \quad
 \mathcal L_0(O):=[H_0,O],
\end{equation}
we have
\begin{equation}
 \mathcal L_0\left(|m\ket\bra n|\right)
 =
 (E_m-E_n)|m\ket\bra n|
 =
 (m-n)|m\ket\bra n|.
\end{equation}
Thus $d=|E_m-E_n|=|m-n|$ is a common distance in the oscillator
matrix basis and the magnitude of a Bohr frequency of the reference
oscillator.  Each shell $d>0$ groups the two frequency sectors $\pm d$
and defines an energy-transfer channel.

On the truncated oscillator space, we use the normalized
Hilbert--Schmidt inner product
$(A|B):=N_{\rm b}^{-1}\operatorname{Tr}_{N_{\rm b}}(A^\dagger B)$.
Let $\Pi_d$ denote the spectral projector of $|\mathcal L_0|$ onto
eigenvalue $d$.  The matrix-element weight in the $d$th shell is
\begin{equation}
 W_d^{(s)}
 :=
 \left.\left(F^{(s)}\right|\Pi_dF^{(s)}\right)
 =
 \frac{1}{N_{\rm b}}
 \sum_{\substack{0\le m,n<N_{\rm b}\\|m-n|=d}}
 \left|F_{mn}^{(s)}\right|^2,
 \label{eq:Wd}
\end{equation}
with $W_d^{(s)}=0$ for odd $d$.  We then define the distance-shell
moments
\begin{align}
 M_p^{(s)}
 &:=
 \sum_{d=0}^{N_{\rm b}-1}d^pW_d^{(s)}
 =
 \left.\left(F^{(s)}\right|\,|\mathcal L_0|^pF^{(s)}\right)
 \nonumber\\
 &=
 \frac{1}{N_{\rm b}}
 \sum_{m,n=0}^{N_{\rm b}-1}
 |m-n|^p\left|F_{mn}^{(s)}\right|^2
 \quad (p\ge1).
 \label{eq:Mp}
\end{align}
Thus, $\{W_d^{(s)}\}$ resolves the reconstructed operator into fixed
energy-transfer shells, while $M_p^{(s)}$ characterizes the distribution
of matrix-element weight across these shells.  This distance-resolved
distribution constitutes the inverse-spectral geometry considered below.

The full construction is summarized by
\begin{align}
\begin{split}
 \{\varepsilon_n^{(s)}\}
 &\xrightarrow{\ \mathcal D\ }
 f^{(s)}(x)
 \xrightarrow{\ \bra m|\cdot|n\ket\ }
 F_{mn}^{(s)}\\
 &\qquad\qquad
 \xrightarrow{\ |m-n|=d\ }
 W_d^{(s)}
 \xrightarrow{\ \sum_d d^p(\cdot)\ }
 M_p^{(s)} .
\end{split}
\end{align}
Here $\mathcal D$ denotes the dressing inverse-spectral map.

For even $p=2q$, \eqref{eq:Mp} reduces to repeated-commutator norms,
\begin{equation}
 M_{2q}^{(s)}
 =
 \frac{1}{N_{\rm b}}
 \left\|\mathcal L_0^qF^{(s)}\right\|_{\rm HS}^2
 \quad (q\ge1),
 \label{eq:M2q}
\end{equation}
so in particular
$M_2^{(s)}=N_{\rm b}^{-1}\|[H_0,F^{(s)}]\|_{\rm HS}^2$.
Thus the even moments form a hierarchy of repeated-commutator norms.

For comparison between input spectra, we use the GUE-normalized ratios
\begin{equation}
 R_p^{(s/{\rm GUE})}
 :=
 \frac{\left\bra M_p^{(s)}\right\ket_s}
 {\left\bra M_p^{({\rm GUE})}\right\ket_{\rm GUE}},
 \label{eq:Rp}
\end{equation}
where $\bra\cdots\ket_s$ denotes an ensemble average for
random-matrix spectra and a window average for the Riemann zeros.
By construction, $R_p^{({\rm GUE}/{\rm GUE})}=1$.

\emph{Dyson universality in the reconstructed operator.---}

\begin{figure*}[t]
 \centering
 \begin{minipage}[t]{0.49\textwidth}
  \centering
  \includegraphics[width=\linewidth]{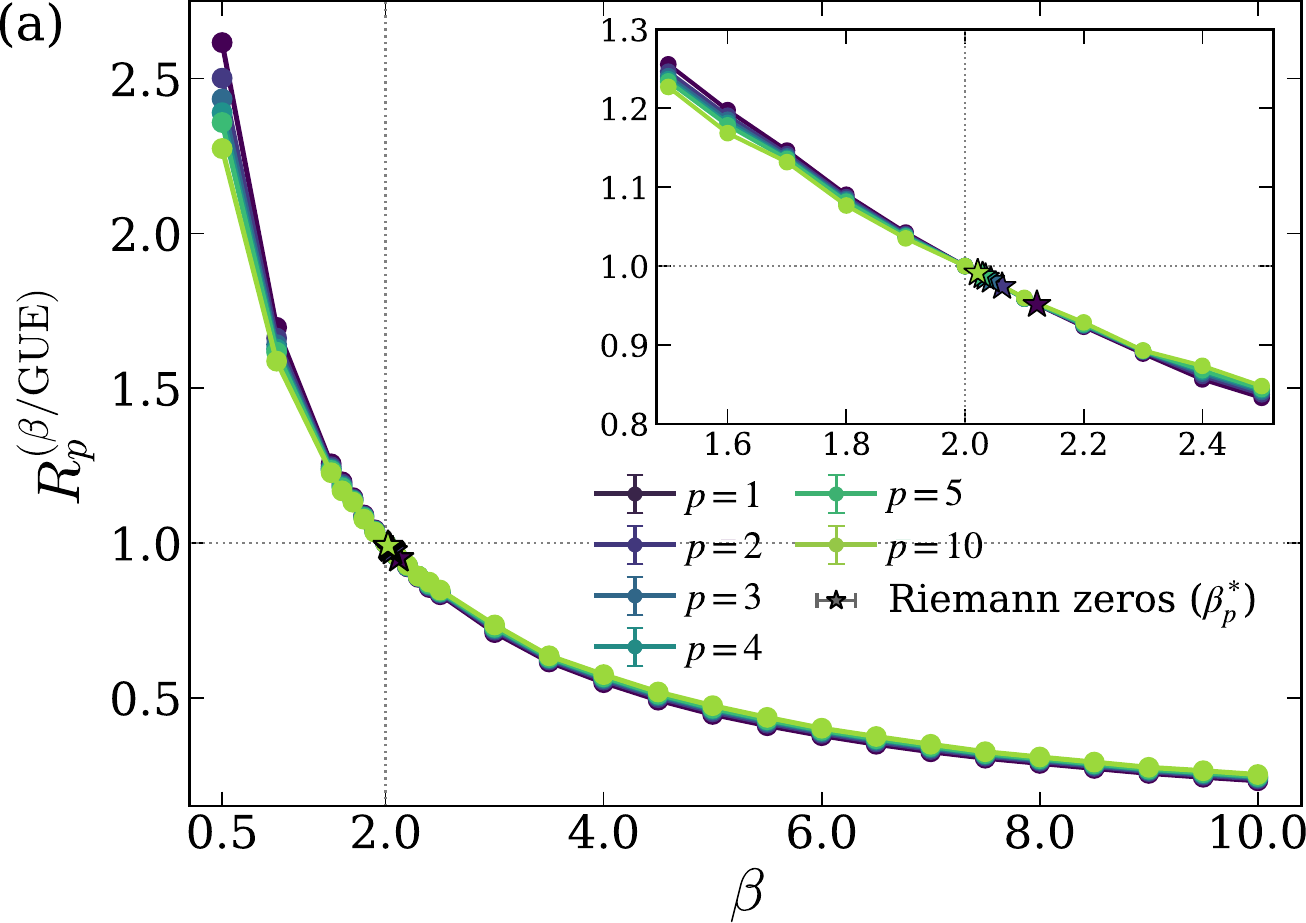}
 \end{minipage}
 \hfill
 \begin{minipage}[t]{0.49\textwidth}
  \centering
  \includegraphics[width=\linewidth]{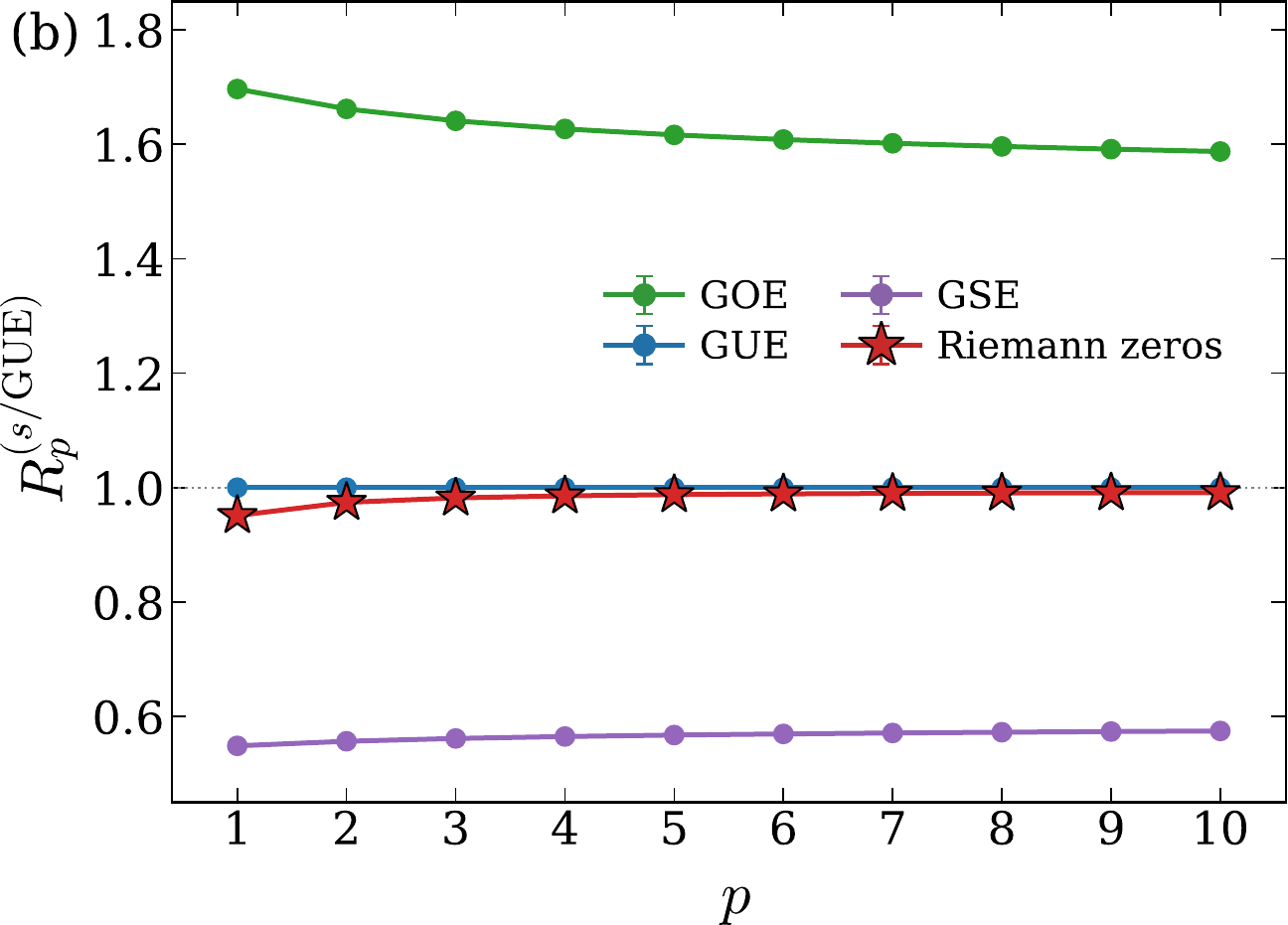}
 \end{minipage}
 \caption{
 Dyson universality in the reconstructed operator.
 (a) GUE-normalized ratios $R_p^{(\beta/{\rm GUE})}$ as functions of
 $\beta$; stars mark the effective indices $\beta_p^\ast$ obtained
 by piecewise-linear inversion at the high-height Riemann-zero ratios.
 (b) $R_p^{(s/{\rm GUE})}$ as functions of $p$ for the GOE, GUE, GSE,
 and Riemann-zero inputs.  Error bars denote bootstrap uncertainties.
 }
 \label{fig:calibration}
\end{figure*}
We first calibrate the GUE-normalized ratios \eqref{eq:Rp} on Gaussian
$\beta$-ensembles, for which the Dyson index is continuously tunable.
The eigenvalues are generated using the Dumitriu--Edelman tridiagonal
matrix model~\cite{DumitriuEdelman2002}.  For each realization, we
retain the central $60\%$ of the eigenvalues to suppress edge effects,
unfold the resulting bulk sequence to unit mean spacing using the
semicircle law, and take the $N_{\rm lev}+1$ consecutive levels
required by the dressing prescription.  No $\beta$-dependent operation
is introduced after unfolding.

The Riemann-zero data used throughout are taken from the LMFDB
\cite{LMFDB}.  The high-height comparison in
Fig.~\ref{fig:calibration} uses $100$ consecutive, nonoverlapping
windows at $T=3.061\times10^{10}$, each containing
$N_{\rm lev}+1$ zeros.  The height study uses $40$ logarithmically
spaced bins over
$1.592\times10^6\le T\le3.061\times10^{10}$, each containing $16$
consecutive, nonoverlapping windows of the same size.  Each window is
unfolded separately to unit mean spacing using the smooth
Riemann--von Mangoldt counting function, and the coordinate $T$ of a
height bin is the center of the total ordinate interval spanned by its
windows.

Unless otherwise stated, we use
$N_{\rm lev}=N_{\rm b}=5\times10^4$, with dressing-grid spacing
$h_x=10^{-4}$ and quadrature spacing $h_{\rm q}=0.005$.  We evaluate
$1\le p\le10$.  The points at $\beta=1,2,4$, the data in
Fig.~\ref{fig:calibration}(b), and the common GUE denominator use
$100$ realizations or windows; the remaining $\beta$ values use $10$.
Error bars are obtained by bootstrap resampling.  Numerical robustness
with respect to $h_x$, $h_{\rm q}$, $N_{\rm b}$, and $N_{\rm lev}$ is
documented in the End Matter.

Fig.~\ref{fig:calibration}(a) shows a smooth response to the continuous
Dyson index, with $R_p^{(\beta/{\rm GUE})}=1$ at $\beta=2$ by
normalization.  Thus, the shell moments resolve the continuous Dyson
index rather than merely separating the three classical Gaussian
ensembles.

With the calibration fixed independently of the zeta data, we define
$\beta_p^\ast$ by
$R_p^{(\beta_p^\ast/{\rm GUE})}=R_p^{(\zeta/{\rm GUE})}$.
Piecewise-linear inversion gives $\beta_1^\ast=2.121(4)$, while
$\beta_p^\ast$ decreases from $2.063(4)$ at $p=2$ to $2.022(5)$ at
$p=10$.  The Riemann-zero values therefore lie close to the GUE point
$\beta=2$, with the largest residual at the lowest moment order.

Fig.~\ref{fig:calibration}(b) shows that the GOE, GUE, and GSE remain
clearly separated throughout $1\le p\le10$, while the high-height
Riemann-zero ratios remain close to GUE.  Since increasing $p$
emphasizes larger Bohr-frequency distances, this moment-order
dependence anticipates the shell-resolved structure discussed below.
By contrast, the coordinate-space box-counting analysis does not
robustly distinguish the GOE, GUE, GSE, and Riemann-zero cases
(see the End Matter).

\emph{Riemann zeros and height dependence.---}
For the Riemann-zero bins defined above, we examine how the
inverse-spectral GUE signature evolves with height.  In the standard
Montgomery--Odlyzko picture, the unfolded zeros approach universal GUE
behavior at high height, while finite-height and longer-range statistics
retain nonuniversal arithmetic corrections associated with primes and
prime powers
\cite{BerryKeating1999,BogomolnyEtAl2006,LugarEtAl2023}.  Holding the
Gaussian $\beta$-ensemble calibration fixed therefore provides a direct
test of whether the inverse-spectral residual from GUE weakens with
height.

Let $\bra\cdots\ket_{\zeta,T}$ denote the average over the $16$
Riemann-zero windows in the bin centered at $T$.  We define
\begin{equation}
 R_p^{(\zeta/{\rm GUE})}(T)
 :=
 \frac{\left\bra M_p^{(\zeta)}(T)\right\ket_{\zeta,T}}
 {\left\bra M_p^{({\rm GUE})}\right\ket_{\rm GUE}},
 \label{eq:Rp-zeta-height}
\end{equation}
using the same independently determined GUE denominator as in the
calibration.

\begin{figure}[t]
 \centering
 \includegraphics[width=\linewidth]{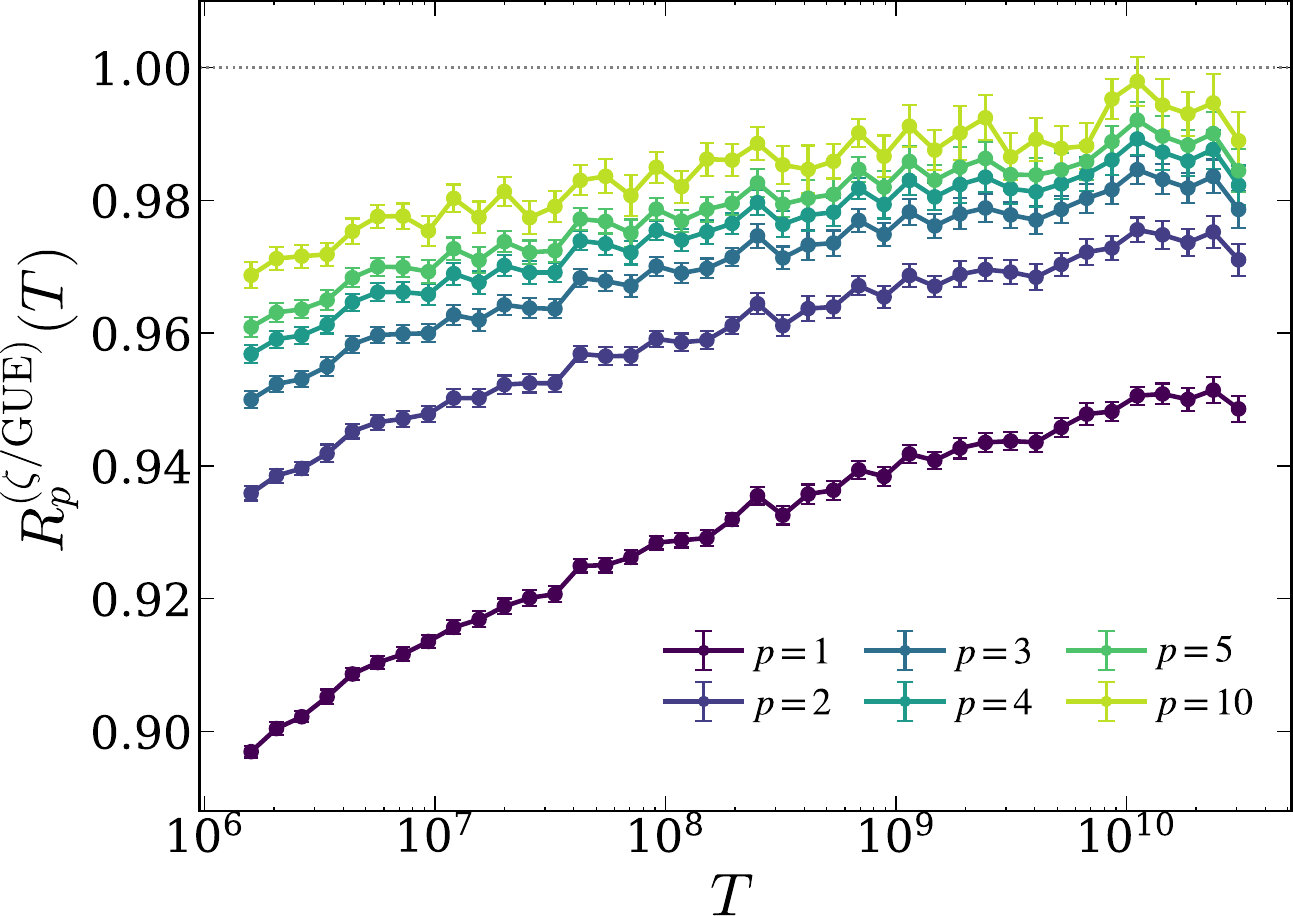}
 \caption{
 Height dependence of the inverse-spectral GUE signature
 $R_p^{(\zeta/{\rm GUE})}(T)$.  Each point averages $16$ consecutive,
 nonoverlapping windows, with $T$ defined as the center of their total
 ordinate interval.  The horizontal dotted line marks the GUE
 normalization.  Error bars denote bootstrap uncertainties.
 }
 \label{fig:zeta-height}
\end{figure}

The height-dependent ratios \eqref{eq:Rp-zeta-height}, shown in
Fig.~\ref{fig:zeta-height}, move overall toward unity over more than
four decades in height, despite fluctuations among individual bins.
The same Gaussian $\beta$-ensemble calibration and GUE denominator are
used throughout,
so the operator-level GUE signature becomes progressively sharper with
height without recalibration.

We do not impose a specific asymptotic law, since the accessible range
does not reliably distinguish among possible scaling forms.  The robust
result is that the inverse-spectral residual weakens with height,
consistent with finite-height arithmetic corrections.  Relating this
residual quantitatively to explicit prime and prime-power contributions
remains an open problem.  Its moment-order dependence suggests that the
residual is distributed nonuniformly over energy-transfer distances,
which we now resolve shell by shell.

\emph{Shell-resolved geometry.---}
To locate the residual in operator space, we resolve the shell weights
\eqref{eq:Wd}, equivalently the spectral measure of $|\mathcal L_0|$,
shell by shell.  For each nonzero even shell, define
\begin{align}
 &\mathcal R_d^{(s/{\rm GUE})}
 :=
 \frac{\bra W_d^{(s)}\ket_s}
 {\bra W_d^{({\rm GUE})}\ket_{\rm GUE}},
 \nonumber\\
 &\mu_p(d)
 :=
 \frac{
 d^p\bra W_d^{({\rm GUE})}\ket_{\rm GUE}
 }{
 \bra M_p^{({\rm GUE})}\ket_{\rm GUE}
 }.
\end{align}
The quantity $\mu_p(d)$ is the normalized contribution of the $d$th shell
to the GUE moment.  Since $\mu_p(d)\ge0$ and
$\sum_d\mu_p(d)=1$ over the nonzero even shells,
\begin{equation}
 R_p^{(s/{\rm GUE})}
 =
 \sum_d
 \mu_p(d)\mathcal R_d^{(s/{\rm GUE})}.
 \label{eq:Rp-shell}
\end{equation}
Thus, $R_p^{(s/{\rm GUE})}$ is a convex combination of the
shell-resolved ratios, with weights $\mu_p(d)$.  For the Riemann zeros,
\eqref{eq:Rp-shell} gives
\begin{equation}
 R_p^{(\zeta/{\rm GUE})}(T)-1
 =
 \sum_d\mu_p(d)
 \left[
 \mathcal R_d^{(\zeta/{\rm GUE})}(T)-1
 \right].
\end{equation}
Hence the deviation of $R_p^{(\zeta/{\rm GUE})}(T)$ from unity is the
$\mu_p$-weighted average of the shell-resolved Riemann--GUE residual.
Increasing $p$ shifts $\mu_p(d)$ toward larger $d$, directly relating
the moment-order dependence to where the residual is localized in the
Liouvillian shells.

\begin{figure}[t]
 \centering
 \includegraphics[width=\linewidth]{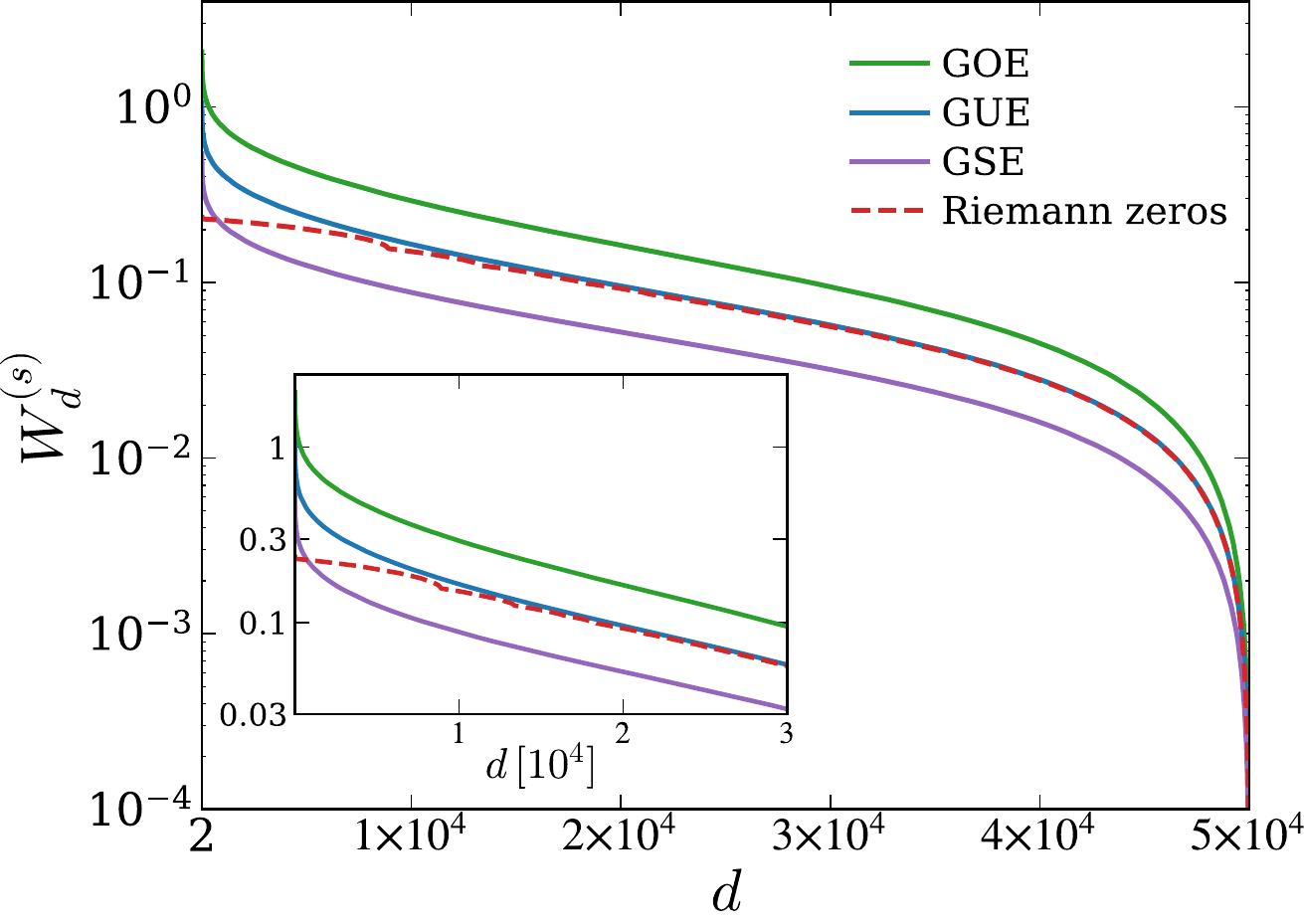}
 \caption{
 Averaged Liouvillian shell weights
 $\bra W_d^{(s)}\ket_s$ for the GOE, GUE, GSE, and the highest-height
 Riemann-zero sample.  The main panel uses a logarithmic vertical scale;
 the inset shows the low-$d$ region on a linear scale.  Only nonzero even
 shells are shown, and $d=0$ is omitted.
 }
 \label{fig:shell-residual}
\end{figure}

Fig.~\ref{fig:shell-residual} shows that the Riemann--GUE residual is
concentrated predominantly in the low-$d$, near-diagonal sectors, while
their shell weights nearly coincide over a broad range of larger
energy-transfer distances.  The approach of
$R_p^{(\zeta/{\rm GUE})}$ to unity therefore reflects a direct matching
of the shell measure rather than an accidental cancellation among shells.
The GOE and GSE retain distinct profiles over the same range.

Because increasing $p$ shifts $\mu_p(d)$ toward the large-$d$ region,
where the Riemann and GUE profiles nearly coincide, the shell-resolved
picture explains the observed moment-order dependence.  Whether the
low-frequency residual is the inverse-spectral image of finite-height
arithmetic corrections associated with primes and prime powers remains
an open question.

\emph{Discussion.---}
The central result is that Dyson universality, ordinarily diagnosed
directly from eigenvalue statistics, reappears after nonlinear
inverse-spectral reconstruction as a geometry of the reconstructed
operator.  The continuous $\beta$ response shows that the shell moments
resolve the Dyson index itself, while for the Riemann zeros the
shell-resolved measure approaches the GUE profile over a broad range of
energy-transfer sectors.  The remaining finite-height deviation is
concentrated predominantly at low frequencies.  The effective indices
$\beta_p^\ast$ therefore provide only moment-dependent projections of a
richer shell-resolved geometry.

This geometry also provides a direct connection to Krylov dynamics.
For Hermitian $F^{(s)}$, the weights $\{W_d^{(s)}\}$ are the spectral
measure of the reference Liouvillian $\mathcal L_0$, folded onto
$d=|\omega|$, and \eqref{eq:M2q} gives its even moments.  In finite
dimension, the complete moment sequence determines the associated
Lanczos coefficients through the Stieltjes construction
\cite{Gautschi2004} and hence the Krylov chain.  Recent work has related
random-matrix universality to
Lanczos spectra and operator growth
\cite{BalasubramanianEtAl2025,SadhasivamEtAl2026}; here the starting
point is reversed, since $F^{(s)}$ is reconstructed from the prescribed
spectrum rather than given a priori.  A more intrinsic extension would
replace the common reference Liouvillian $\mathcal L_0$ by
$\mathcal L_s=\operatorname{ad}_{H^{(s)}}$.  Identifying the arithmetic
origin of the low-frequency residual and extending the construction to
other symmetry families of $L$-functions are further natural problems.

\emph{Acknowledgments.---}
This work was supported by JSPS KAKENHI Grant No.~JP24K06889.

\paragraph*{Data and code availability.---}
The code and processed data used to generate the results and figures in
this work are publicly available at
\url{https://github.com/momohayashi304/Riemann_Inverse-Spectral_Geometry}.

\bibliographystyle{apsrev4-2}
\bibliography{BibFile}

\section*{END MATTER}

\emph{Dressing transformation, energy restoration, and parity.---}

\begin{figure*}[ttt]
 \centering
 \includegraphics[width=\textwidth]{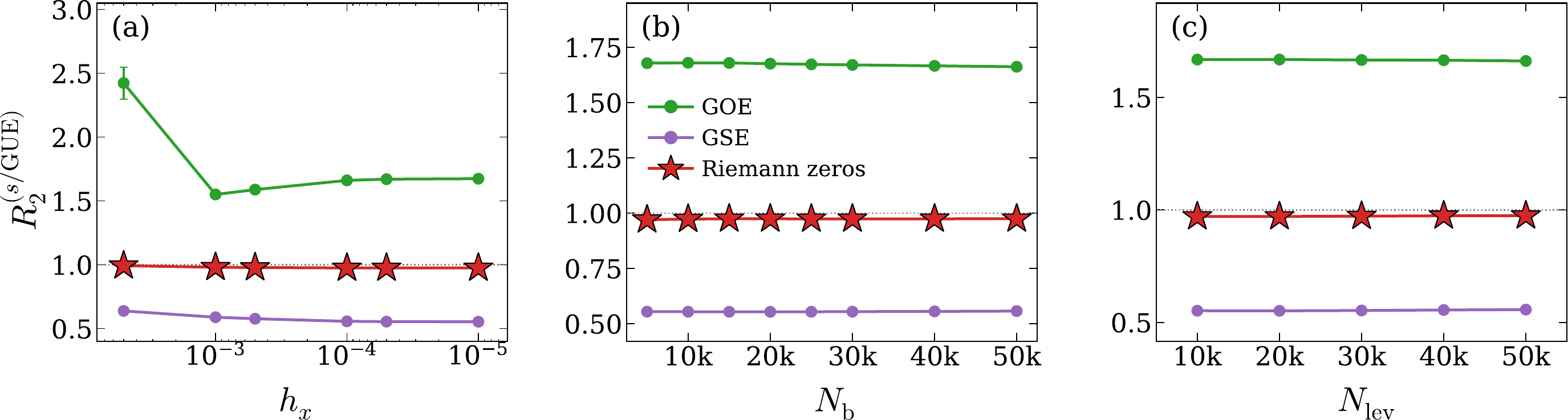}
 \caption{
 Numerical robustness of the representative ratio
 $R_2^{(s/{\rm GUE})}$ for the GOE, GSE, and Riemann-zero inputs.
 (a) Dependence on the dressing-grid spacing $h_x$ at fixed
 $N_{\rm b}$ and $N_{\rm lev}$.
 (b) Dependence on the oscillator-basis size $N_{\rm b}$ at fixed
 $h_x$ and $N_{\rm lev}$.
 (c) Joint finite-size scan with $N_{\rm b}=N_{\rm lev}$ at fixed
 $h_x$.  Panel (a) uses $50$ realizations or windows, whereas panels
 (b) and (c) use $100$.  Error bars denote bootstrap uncertainties.
 }
 \label{fig:numerical-robustness}
\end{figure*}

We briefly describe the inverse-spectral map used in the main text.
The unfolded target levels
$\{\varepsilon_n^{(s)}\}_{n=0}^{N_{\rm lev}-1}$ are supplemented by an
auxiliary upper level $\varepsilon_{N_{\rm lev}}^{(s)}$, and we define
\begin{equation}
 \varepsilon_{\rm sh}^{(s)}
 :=
 \varepsilon_{N_{\rm lev}}^{(s)},
 \quad
 \widetilde{\varepsilon}_j^{(s)}
 :=
 \varepsilon_j^{(s)}-\varepsilon_{\rm sh}^{(s)}
 \quad
 (0\le j\le N_{\rm lev}).
\end{equation}
Thus
$\widetilde{\varepsilon}_0^{(s)}<\cdots<
\widetilde{\varepsilon}_{N_{\rm lev}-1}^{(s)}<0$ and
$\widetilde{\varepsilon}_{N_{\rm lev}}^{(s)}=0$.
The auxiliary level fixes the energy origin of the recursion and is not
inserted as a bound state.

Following \cite{RamaniEtAl1995,vanZylHutchinson2003}, we consider
the Schr\"odinger eigenvalue problems
\begin{equation}
 K_j^{(s)}\psi_n^{(j,s)}
 =
 \widetilde{\varepsilon}_n^{(s)}\psi_n^{(j,s)},
 \quad
 K_j^{(s)}
 :=
 -\frac{d^2}{dx^2}+U_j^{(s)}(x).
\end{equation}
At stage $j$, the prescribed discrete levels are
$\widetilde{\varepsilon}_j^{(s)},\ldots,
\widetilde{\varepsilon}_{N_{\rm lev}-1}^{(s)}$.
The construction starts from the free problem
$K_{N_{\rm lev}}^{(s)}=-d^2/dx^2$, with
$U_{N_{\rm lev}}^{(s)}(x)=0$.

The levels are inserted successively from the top down.  Suppose that
$K_{j+1}^{(s)}$ already contains
$\widetilde{\varepsilon}_{j+1}^{(s)},\ldots,
\widetilde{\varepsilon}_{N_{\rm lev}-1}^{(s)}$.
To insert the next lower level $\widetilde{\varepsilon}_j^{(s)}$, we
choose an even, nodeless solution of
\begin{equation}
 \left[
 -\frac{d^2}{dx^2}+U_{j+1}^{(s)}(x)
 \right]
 \varphi_j^{(s)}(x)
 =
 \widetilde{\varepsilon}_j^{(s)}
 \varphi_j^{(s)}(x).
 \label{eq:dressing-seed}
\end{equation}
Since $\widetilde{\varepsilon}_j^{(s)}$ lies below the bottom of the
spectrum of $K_{j+1}^{(s)}$, the seed can be chosen without nodes.
Nodelessness prevents singularities in the dressed potential.

Introducing
$w_j^{(s)}:=-d\log\varphi_j^{(s)}/dx$, the seed equation
\eqref{eq:dressing-seed} becomes the Riccati equation
\begin{equation}
 \frac{dw_j^{(s)}}{dx}
 -
 \left(w_j^{(s)}\right)^2
 +
 U_{j+1}^{(s)}(x)
 =
 \widetilde{\varepsilon}_j^{(s)},
\end{equation}
which is integrated numerically.  The transformed potential is
\begin{equation}
 U_j^{(s)}(x)
 =
 U_{j+1}^{(s)}(x)
 -
 2\frac{d^2}{dx^2}\log\varphi_j^{(s)}(x).
 \label{eq:dressing-step}
\end{equation}

Equivalently, with
$A_j^{(s)}:=d/dx+w_j^{(s)}(x)$ and
$A_j^{(s)\dagger}:=-d/dx+w_j^{(s)}(x)$, the Darboux factorization reads
\begin{equation}
 K_{j+1}^{(s)}-\widetilde{\varepsilon}_j^{(s)}
 =
 A_j^{(s)\dagger}A_j^{(s)},
 \quad
 K_j^{(s)}-\widetilde{\varepsilon}_j^{(s)}
 =
 A_j^{(s)}A_j^{(s)\dagger}.
\end{equation}
The chosen seed is non-normalizable, while its reciprocal is
normalizable.  Hence $K_j^{(s)}$ acquires a new bound state proportional
to $1/\varphi_j^{(s)}$ at $\widetilde{\varepsilon}_j^{(s)}$, while the
previously inserted levels are preserved by the Darboux intertwining.
Thus each step inserts precisely one new lowest level.  Iterating from
$j=N_{\rm lev}-1$ to $j=0$ therefore produces $K_0^{(s)}$ with discrete
spectrum
$\{\widetilde{\varepsilon}_n^{(s)}\}_{n=0}^{N_{\rm lev}-1}$.

Restoring the original energy origin gives
\begin{equation}
 H^{(s)}
 :=
 K_0^{(s)}+\varepsilon_{\rm sh}^{(s)}
 =
 -\frac{d^2}{dx^2}
 +
 U_0^{(s)}(x)
 +
 \varepsilon_{\rm sh}^{(s)},
\end{equation}
whose prescribed low-lying eigenvalues are
$\{\varepsilon_n^{(s)}\}_{n=0}^{N_{\rm lev}-1}$.
Relative to the reference oscillator
$H_0=-d^2/dx^2+x^2/4$, the reconstructed deformation is
\begin{equation}
 f^{(s)}(x)
 :=
 U_0^{(s)}(x)
 +
 \varepsilon_{\rm sh}^{(s)}
 -
 \frac{x^2}{4},
\end{equation}
so that $H^{(s)}=H_0+f^{(s)}(x)$.

Finally, parity is preserved throughout the chain.  The initial
potential is even, and an even seed is chosen at every step.
Eq.~\eqref{eq:dressing-step} therefore preserves evenness, so
$U_j^{(s)}(-x)=U_j^{(s)}(x)$ and hence
$f^{(s)}(-x)=f^{(s)}(x)$.  Together with the parity $(-1)^m$ of the
oscillator eigenfunctions, this gives
$F_{mn}^{(s)}=0$ for odd $|m-n|$, as used in the main text.
The same dressing prescription is applied to every Gaussian
$\beta$-ensemble realization and every Riemann-zero window.

\emph{Numerical implementation and robustness.---}

All input spectra are processed with the same dressing prescription and
oscillator projection.  The matrix elements \eqref{eq:Fmn} are
evaluated over $|x|\le x_{\rm cut}$, where
\begin{equation}
 x_{\rm cut}
 :=
 \sqrt{4(N_{\rm b}-1)+2}+15
\end{equation}
is the classical turning point of the highest retained oscillator state
plus a fixed margin.  The production parameters are
$h_x=10^{-4}$, $h_{\rm q}=0.005$, and
$N_{\rm b}=N_{\rm lev}=5\times10^4$, as specified in the main text.
Statistical uncertainties are estimated by bootstrap resampling of the
independent realizations or nonoverlapping Riemann-zero windows.

The restriction $p\le10$ controls sensitivity to the finite basis edge.
For an allowed shell $d>0$, there are $2(N_{\rm b}-d)$ matrix elements.
A geometric shell-counting estimate therefore gives, for large
$N_{\rm b}$, the fraction of $M_p$ arising from
$d>\alpha N_{\rm b}$ as
\begin{equation}
 \eta_p(\alpha)
 \simeq
 \frac{\displaystyle\int_\alpha^1 x^p(1-x)\,dx}
 {\displaystyle\int_0^1 x^p(1-x)\,dx}
 =
 1-(p+2)\alpha^{p+1}+(p+1)\alpha^{p+2}.
 \label{eq:edge-estimate}
\end{equation}
For $\alpha=0.9$, \eqref{eq:edge-estimate} gives $5.23\%$, $34.10\%$,
and $66.08\%$ for $p=2,10$, and $20$, respectively, in close agreement
with the numerical GUE fractions $4.90\%$, $34.18\%$, and $66.15\%$.
Thus $p\le10$ retains sensitivity to large energy-transfer distances
without allowing the basis edge to dominate.

Fig.~\ref{fig:numerical-robustness} shows that the Dyson-class separation
and the placement of the Riemann-zero data near GUE are stable under
variations of $h_x$, $N_{\rm b}$, and $N_{\rm lev}$.  The small residual
drifts are well below the separation between the Dyson classes and do
not alter the GUE assignment.  Representative scans at
$p=1,3,4,5$, and $10$ give the same class-level conclusion as the
$p=2$ results shown in Fig.~\ref{fig:numerical-robustness}.

The quadrature discretization entering the matrix elements was checked
separately.  Halving the production spacing from $h_{\rm q}=0.005$ to
$h_{\rm q}=0.0025$ produces no appreciable change in
$R_p^{(s/{\rm GUE})}$ for the GOE, GSE, and Riemann-zero inputs over
$1\le p\le10$.  As a complementary consistency check, the
high-resolution box-counting analysis below shows that the reconstructed
potentials cross over from their intermediate-scale roughness near
$\ell\sim10^{-2}$, above the production quadrature spacing.  These
results support the use of $h_{\rm q}=0.005$ in evaluating
\eqref{eq:Fmn}.

These tests establish numerical stability of the normalized ratios;
the raw moments $M_p^{(s)}$ may retain a common dependence on the finite
basis cutoff that cancels in $R_p^{(s/{\rm GUE})}$.

\emph{Box-counting control.---}
We finally test whether the Dyson-class dependence is already visible
in the coordinate-space roughness of the reconstructed potential
$f^{(s)}(x)$.  This provides a natural control, since inverse-spectral
potentials associated with the Riemann zeros have traditionally been
characterized by their fractal structure
\cite{WuSprung1993,vanZylHutchinson2003,
SchumayerEtAl2008,SchumayerHutchinson2011}.

Let $N_{\rm box}^{(s)}(\ell)$ denote the number of boxes of side length
$\ell$ required to cover the graph of $f^{(s)}(x)$.  The production
deformations, generated with the dressing spacing $h_x=10^{-4}$, are
analyzed on the common central interval $x\in[0,10]$.  The box-counting
analysis is performed directly on this high-resolution grid, allowing
the small-$\ell$ crossover to be resolved.  We extract an effective
intermediate-scale box-counting exponent by fitting
$\log N_{\rm box}^{(s)}(\ell)$ versus $\log(1/\ell)$ over
$\ell\in[0.025,0.5]$.

\begin{figure}[b]
 \centering
 \includegraphics[width=\linewidth]{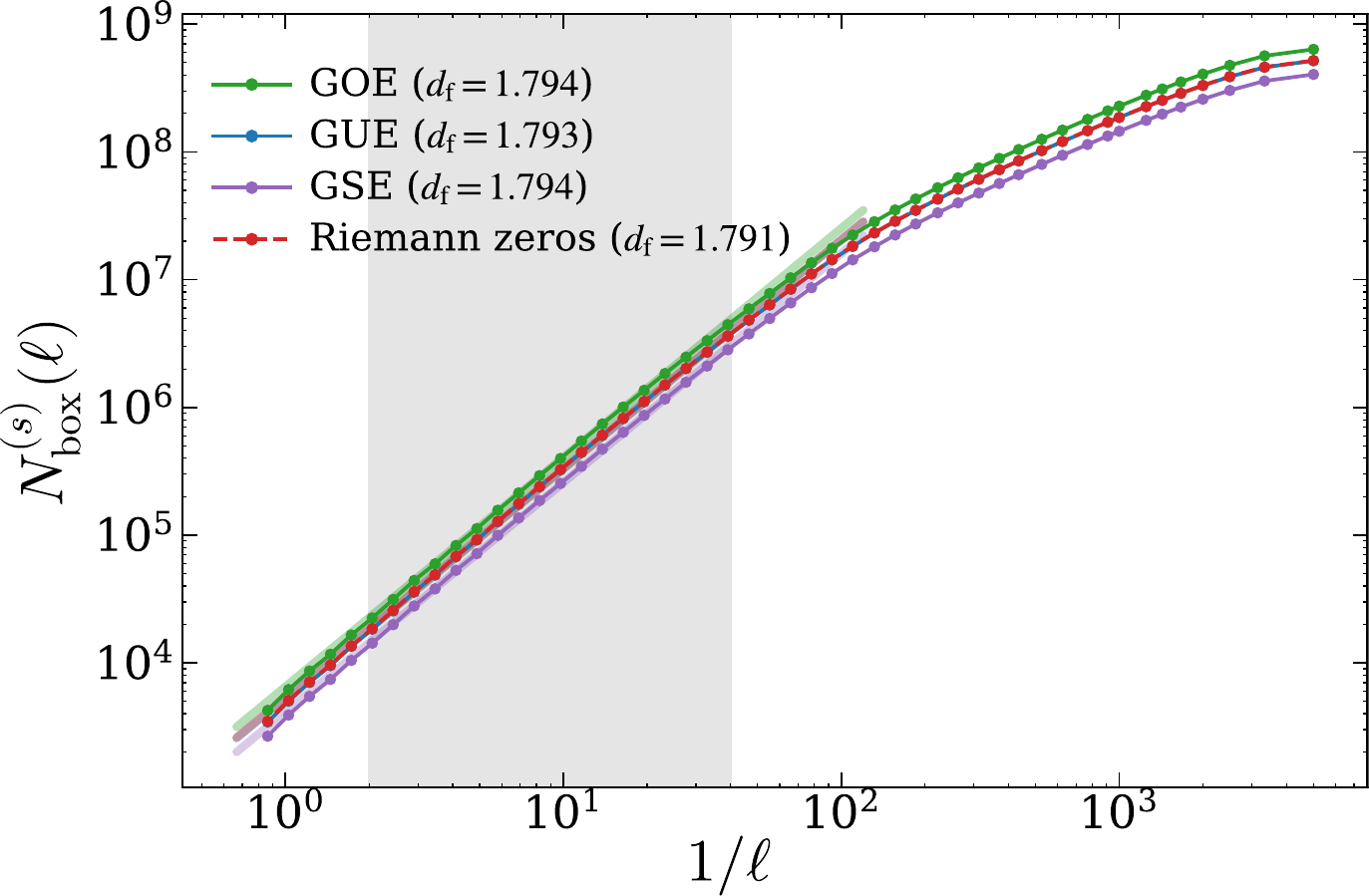}
 \caption{
 Box-counting control for the reconstructed deformation potentials.
 Mean box counts $N_{\rm box}^{(s)}(\ell)$ for the GOE, GUE, GSE,
 and Riemann-zero inputs on the common interval $x\in[0,10]$.
 The shaded region indicates the fitting range
 $\ell\in[0.025,0.5]$, yielding effective intermediate-scale
 exponents $d_f\simeq1.79$ for all four cases.  The curves exhibit
 a common crossover near $\ell\sim10^{-2}$, with no robust
 Dyson-class separation.
 }
 \label{fig:fractal-dimension}
\end{figure}

As shown in Fig.~\ref{fig:fractal-dimension}, the four cases exhibit
essentially the same intermediate-scale box-counting behavior, with
effective exponents $d_f\simeq1.79$.  At smaller scales, the log--log
curves cross over near $\ell\sim10^{-2}$ rather than continuing with a
scale-independent slope.  The Dyson-class dependence is therefore not
resolved by the coordinate-space roughness of $f^{(s)}(x)$; it emerges
only after the deformation is represented in the common oscillator
basis and its matrix-element weight is resolved into energy-transfer
shells.

The location of this crossover provides an additional consistency check
on the quadrature used for the matrix elements: the production spacing
$h_{\rm q}=0.005$ is finer than the observed crossover scale.
\end{document}